# Review on methods of solving the refractive index–thickness coupling problem in digital holographic microscopy of biological cells

Gili Dardikman and Natan T. Shaked[*]

Department of Biomedical Engineering, Faculty of Engineering, Tel Aviv University, Tel Aviv 69978, Israel

*Corresponding author: nshaked@tau.ac.il

**Abstract**

Digital holographic microscopy is a thriving imaging modality that attracted considerable research interest in quantitative biological cell imaging due to its ability to not only create excellent label-free contrast, but also supply valuable physical information regarding the density and dimensions of the sample with nanometer-scale axial sensitivity. This technique records the interference pattern between a sample beam and a reference beam, and by digitally processing it, one can reconstruct the optical path delay between these beams. Per each spatial point, the optical path delay map is proportional to the product of the sample physical thickness and the integral refractive index of the sample. Since the refractive index of the cell indicates its contents without the need for labeling, it is highly beneficial to decouple cell physical thickness from its refractive index profile. This manuscript reviews various approaches of extracting the refractive index from digital holographic microscopy measurements of cells. As soon as the refractive index of the cell is available, it can be used for either biological assays or medical diagnosis, as reviewed in this manuscript.

## 1. Introduction

Imaging live biological cells in vitro is of great importance for both biological research and clinical diagnostics. Yet, isolated cells in vitro have very low amplitude modulation, causing standard amplitude-based imaging (bright-field microscopy) to have poor contrast. Cell staining or labeling is often used to obtain better contrast, yet it is time consuming, sometimes suffers from photobleaching, and may disturb the cellular behavior of interest [1]. The phase profile of the sample is a part of its complex wave front that encodes how much light was delayed when interacting with the sample, and is proportional to the product of the cell thickness and the average refractive index (RI), implying on the cell local density. Since different cellular organelles have different densities and geometries, phase encompasses excellent label-free contrast potential. While in conventional imaging phase cannot be captured due to the lengthy detector integration time relative to the speed of light, digital holographic microscopy (DHM) captures the phase difference between a beam that interacted with the sample (typically passes through it for cells in vitro) and a beam that did not (reference beam); this is done by recording their interference pattern (digital hologram) created on the digital camera, thus converting the phase into intensity variations that can be recorded by the camera [2, 3]. This method holds great promise, as the phase delay does not only supply good contrast, but also consists of valuable information regarding both the thickness of the sample in the direction of light propagation and the RI distribution of the sample on each spatial point on the sample, and thus is considered a quantitative imaging method (in contrast to Zernike's phase contrast and differential interference contrast (DIC) microscopies) [4]. The cell RI profile holds great potential for medical diagnosis and for biological research, since it is a physical measurement of the contents of the sample. Nevertheless, after retrieving the quantitative phase profile from the recorded digital hologram, the geometrical thickness and RI information are coupled in a way that makes it difficult to decipher each of these properties separately. For example, when applying hypotonic shock to cells, their swelling is characterized by a thickness increase combined with RI decrease as a result of dilution; in the absence of a decoupling strategy, though, cellular swelling is typically measured as a phase decrease, which may often be inaccurate [5].

Several methods have been suggested for dealing with the refractive index-thickness coupling problem in DHM. These methods are reviewed in this paper.

The first and most direct one is not trying to solve the coupling problem at all, but rather directly isolating unique characterizing parameters such as dry mass, cell area or frequency content, based on the quantitative phase values themselves, enabling classification based on the raw phase images [4, 6-16]. An equally simple method is relying on exiting RI statistics for cell organelles given in the literature to retrieve the cell physical thickness; this is particularly useful for homogenous cells, such as red blood cells, where the RI is uniform [17-20].

Another group of methods solves the RI-thickness coupling problem by evaluating the thickness of the sample at each spatial location, which allows the isolation of the average RI in that pixel (also called integral RI). The simplest and fastest method is approximating the local thickness based on the fact that cells in suspension assume a spherical shape [21-28], yielding the integral RI 2-D profile of the cell from its quantitative phase profile with no prior knowledge other than the RI of the suspension medium. Another option is using a different imaging method to directly measure the geometrical thickness, enabling the isolation of the integral RI [29-32]. A third option is not measuring the native thickness of the sample but rather constraining the cells into a known dimensional microstructure that confines the cell in the vertical direction such that its thickness is known [21].

Another approach is performing two interferometric measurements, each of them with a different surrounding medium [5, 33, 34] or a different wavelength [35-37], thus retrieving two phase profiles, yielding two linear equations with two unknowns for each spatial location, enabling decoupling of the integral RI from the thickness.

The third approach is tomographic phase microscopy (TPM); this method enables not only to decouple the cell thickness and the integral 2-D RI profile, but rather to obtain the 3-D distribution of the RI of the cell. This is achieved by capturing phase images of the sample from multiple viewing angles, and digitally processing all of them to the 3-D RI index distribution [38-56].

This manuscript is constructed as follows. First, in Section 2, we explain the theory of the RI-thickness coupling problem. Then, in Section 3 we review decoupling methods involving the extraction of the integral 2-D RI by thickness evaluation, either by approximation, direct measurement, or confinement. In Section 4, we analyze methods solving the coupling problem by preforming two different interferometric measurements, yielding two equations with two unknowns. In Section 5, we review setups and algorithms for reconstructing the 3-D RI. Afterwards, in Section 6, we review medical and biological applications for which the RI measurement is useful. Finally, Section 7 concludes this review.

## 2. Theory of the RI-thickness coupling problem

The phase difference, $\varphi$, between the sample and reference wave is proportional to the optical path difference (OPD) between these beams, as following:

$$\varphi(x,y) = \frac{2\pi}{\lambda} \cdot \text{OPD}(x,y), \tag{1}$$

where $\lambda$ is the illumination wavelength. Neglecting diffraction for simplicity, the OPD can be written as:

$$\text{OPD}(x,y) = \int_0^{h(x,y)} [n(x,y,z) - n_m] dz, \tag{2}$$

where $z$ is the direction of light propagation, $h(x,y)$ is the thickness of the sample in the $z$ dimension, $n(x,y,z)$ is the RI distribution of the sample, and $n_m$ is the RI of the medium.

In a discrete representation, the OPD can be described as a finite sum:

$$\text{OPD}(p,q) = \sum_{l=1}^{N(p,q)} [n(p,q,l) - n_m] \cdot \Delta l, \tag{3}$$

where $N(p,q)$ is the number of discrete increments of the sample in the $z$ dimension for pixel $(p,q)$, and $\Delta l$ is the discrete increment length in the $l$ dimension, given by:

$$\Delta l = \frac{\Delta_{CCD}}{M}, \tag{4}$$

where $\Delta_{CCD}$ is the pixel size in the digital camera and $M$ is the total optical magnification used in the setup. In Eq. (4), we assume that the $l$ dimension increment size is the same as the $p$ dimension increment size.

Since $\Delta l$ and $n_m$ are constants and do not depend on $l$, we can take them out of the sum. We can also multiply and divide by $N(p,q)$. Altogether, we get:

$$\text{OPD}(p,q) = \Delta l \cdot N(p,q) \cdot \left[\frac{\sum_{l=1}^{N(p,q)} n(p,q,l) - N(p,q) \cdot n_m}{N(p,q)}\right], \tag{5}$$

which is equivalent to:

$$\text{OPD}(p,q) = h(p,q) \cdot \left[\frac{\sum_{l=1}^{N(p,q)} n(p,q,l)}{N(p,q)} - n_m\right], \tag{6}$$

meaning that the OPD in each pixel is the product of the thickness of the sample at that point with the difference between the $l$-dimension average (or integral) RI of the sample in that point and the medium:

$$\text{OPD}(p,q) = h(p,q) \cdot [n_{cell}(p,q) - n_m], \tag{7}$$

where $n_{cell}(p,q)$ is the integral RI distribution of the cell. The OPD by itself is not a conventional phyisical qunatity, thus for many biological and medical assays, one will first need to decouple the thickness of the sample, $h(p,q)$, and the integral RI of the cell, $n_{cell}(p,q)$. This defines the RI-thickness coupling problem, for which the next sections presents solutions.

### 3. Extracting the integral RI by thickness evaluation

If the thickness profile of the cell $h(p,q)$ is given, the integral RI $n_{cell}(p,q)$ can be simply obtained from Eq. (7), as follows:

$$n_{cell}(p,q) = \frac{\text{OPD}(p,q)}{h(p,q)} + n_m , \qquad (8)$$

where $\text{OPD}(p,q)$ is proportional to the phase profile according to Eq. (1) and $n_m$ is assumed to be known (usually 1.33, the RI of water). The next subsections present three approaches, approximation, direct measurement and thickness confinement, for obtaining $h(p,q)$ before the holographic analysis is performed.

*3.1 Extracting the RI by approximated modeling the cell thickness*

Several methods simply approximate the cell in suspension as a sphere [21-27]. As can be seen in Fig. 1(a), the thickness profile of a sphere with radius $R$ centerd at the $(p_0, q_0)$ pixel is given by:

$$h(p,q) = \begin{cases} 2\Delta l \cdot [R^2 - (p-p_0)^2 - (q-q_0)^2]^{0.5} & \text{for} \quad (p-p_0)^2 + (q-q_0)^2 \leq R^2 \\ 0 & \text{for} \quad (p-p_0)^2 + (q-q_0)^2 > R^2 \end{cases}, \qquad (9)$$

where $\Delta l$ is the image scale, that can be calculated using the known setup parameters and Eq. (4).

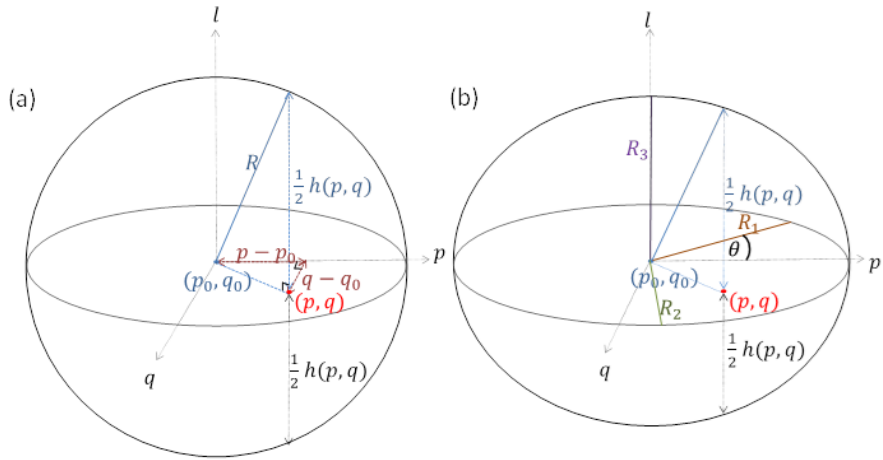

**Fig 1:** Schematic illustrations for various cell-fitting approaches: (a) Calculation of the thickness profile of a sphere with radius $R$ centerd at the $(p_0, q_0)$ pixel. (b) Calculation of the thickness profile of an ellipsoid model. $R_1, R_2, R_3$ are the radii of the cell, and $\theta$ is the orientation of the ellipse. The diameters in the $p$ and $q$ dimensions are the major and minor axes extracted from the OPD image, and the diameter in the $l$ dimension is estimated to be their average. The blue circle denotes the center of mass. The Cartesian coordinate system indicates directions, where light propagates in the $l$ direction and the OPD image is in the $(p,q)$ plain.

There are several approaches for fitting a sphere to the phase map of the cell [21-27]. Kemper et al. [22, 23] suggested to approximate the integral RI profile by applying an iterative fitting of the entire OPD given in Eq.(7) using the Gauss-Newton method for solving non-linear least squares problems [57], where $h(p,q)$ is given by Eq. (9) such that we remain to fit $n_{cell}$, $R$, $p_0$, and $q_0$. In this method, $n_{cell}$ is assumed to be constant, an assumption that is valid for cells with a homogenous RI, where a spherical morphology also dictates a spherical OPD profile.

Other approaches suggest not using the phase or OPD values themselves to fit the sphere, but rather only the morphology, yielding a more general model that can be applied to non-homogenous cells as well; Schürmann et al. [25] suggested obtaining the sphere radius by fitting a circle to the edge of the cell in the phase image, where the edges can be determined with an edge detection algorithm such as Canny edge detection [58]. Steelman et al. [27] suggested fitting the circle using a circular Hough Transform to create a thickness profile [59,60]. In both approaches, once the radius is fitted, Eq. (9) can be used to calculate the thickness profile. We have recently suggested expanding the search for an ellipsoid to account for cases where the cell is not perfectly spherical [28] (see Fig. 1(b)); this can be done by finding the area of the cell using a simple threshold on the phase values, followed by finding the minor and major axes lengths $(2R_1, 2R_2)$ and orientation $(\theta)$ of the ellipse that has the same normalized second central moments as the area, as well as the center of mass $(p_0, q_0)$. Based on the minor and major axes lengths (which should be very similar), the length of the third, orthogonal diameter $(2R_3)$ can be calculated as the average between them. The thickness attributed to the cell at each pixel can then be calculated using the following formula:

$$h(p,q) = \begin{cases} 2\Delta l \cdot R_3 \left[1 - \frac{[(p-p_0)\cos\theta+(q-q_0)\sin\theta]^2}{R_1^2} - \frac{[(p-p_0)\sin\theta-(q-q_0)\cos\theta]^2}{R_2^2}\right]^{0.5}, \\ \qquad \text{for } \frac{[(p-p_0)\cos\theta+(q-q_0)\sin\theta]^2}{R_1^2} + \frac{[(p-p_0)\sin\theta-(q-q_0)\cos\theta]^2}{R_2^2} \leq 1 \\ 0, \qquad \text{for } \frac{[(p-p_0)\cos\theta+(q-q_0)\sin\theta]^2}{R_1^2} + \frac{[(p-p_0)\sin\theta-(q-q_0)\cos\theta]^2}{R_2^2} > 1 \end{cases} \quad (10)$$

Note that when assuming a non-uniform cell RI, points near the edge should be excluded since the thickness at these points approaches zero and the RI becomes ill defined [25].

An example of applying the latter method that was recently suggested by us [28] to a human colorectal adenocarcinoma colon SW480 cell is given in Fig. 2.

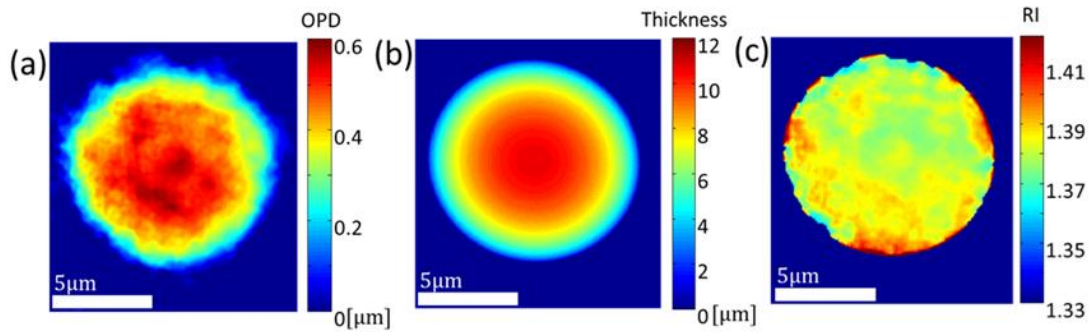

**Fig 2: Result of applying the ellipsoidal fit decoupling technique to an SW480 cell. (a) Measured OPD image, fitted with an ellipse with major axes radii $R_1 = 5.39$ μm and $R_2 = 5.25$ μm. (b) Thickness evaluation, calculated using Eq. (13). (c) Resulting integral RI profile.**

More examples of integral RI profiles received by thickness estimation can be seen in Refs. [24] and [26].

*3.2 Extracting the integral RI by direct thickness measurement*

Both Cardenas et al. [29] and Balberg et al. [30] have previously suggested using an atomic force microscope (AFM) [61] to yield a precise label-free thickness measurement, later utilized for decoupling RI and thickness in the phase measurement. Furthermore, since the transverse resolution of DHM is diffraction limited, co-registration of AFM image provides higher transverse resolution at nanometer scale [29]. Cardenas et al. [29] have integrated the DHM and AFM into a single system allowing for simultaneous measurement, where the intrusion of the AFM probe is minimal during simultaneous AFM and DHM recording due to the transparent nature and bent configuration of the optical fiber-based AFM cantilever. Balberg et al. [30] have used a specialized grid to identify fixated cells imaged using DHM and correlate them with the ones measured using the AFM.

Alternatively, confocal fluorescence microscopy was also previously used to directly measure the physical thickness of the cells [33, 62], yet it usually requires chemical staining such as fluorescence dyes. Lue et al. [31] have suggested using confocal reflectance microscopy [63], which can provide the physical contour of the cell without staining. This enables high resolution cell thickness maps due to its sectioning capability. Then, again, the thickness measurement can be used for solving the thickness-RI coupling problem in DHM.

Choi et al. [32] have suggested full-field optical coherence microscopy to measure the thickness of the sample, by taking a series of 0.6 μm resolved tomograms. For the specific case of samples with reflective surface, variations in thickness can also be quantified by standard reflection mode DHM, enabling quick and precise imaging [34, 64]. Yet, most biological samples are not reflective.

Note that all above techniques are valid only if the thickness of the cell remains unchanged between the two measurements, as would be true for static cells. Even for measurements that are carried out simultaneously, this can be an issue if the height measurement time is lengthy (e.g., includes sample scanning).

Examples of integral RI profiles obtained by a direct thickness measurement are given in Ref. [31].

*3.3 Extracting the integral RI by thickness confinement*

Physically confining the cells to a certain thickness is another approach for solving the RI-thickness coupling problem. Lue et al. [21] have placed live cells in microchannels of fixed dimensions that confine the cell in the axial direction; single input and single output microchannels of rectangular cross sections were prepared by molding elastomer on microstructures fabricated on a silicon wafer [65]. Since the microchannel thickness exhibits some variability due to the fabrication process, a relative measurement can be performed by filling the cell with a medium of a known RI $n_M$, eliminating the need for a priori knowledge of the microchannel constant thickness distribution $h(p,q)$. Given a liquid with RI $n_M$, its OPD ($\text{OPD}_l(p,q)$) is:

$$\text{OPD}_l(p,q) = h(p,q) \cdot [n_M - n_m], \tag{11}$$

enabling the extraction of the thickness distribution of the microchannel (and thus of the confined cell):

$$h(p,q) = \frac{[n_M - n_m]}{\text{OPD}_l(p,q)}. \tag{12}$$

Lue et al. [21] have also suggested that this technique can be automated by combining the imaging geometry with flow in the microfluidic channels, allowing for high-throughput cytorefractometry.

4. **Extracting the integral RI by two different interferometric measurements**

To solve the RI-thickness coupling problem in DHM, it has been proposed to put the cell in a perfusion chamber, while attached to the bottom plane, and record the given biological cell at two different surrounding media with different RIs of a known value by a perfusion of different media. Then, the two extracted OPD profiles create two linear equations with two unknowns for each pixel, $n_{cell}(p,q)$ and $h(p,q)$, enabling the decoupling of the integral RI from the thickness [5,33,34]. According to Eq. (7), the two measurements yield the following two OPD profiles:

$$\begin{cases} \text{OPD}_1(p,q) = h(p,q) \cdot [n_{cell}(p,q) - n_{m,1}] \\ \text{OPD}_2(p,q) = h(p,q) \cdot [n_{cell}(p,q) - n_{m,2}] \end{cases}. \tag{13}$$

where $\text{OPD}_1(p,q)$ is the OPD measured with the surrounding medium that has RI, $n_{m,1}$, and $\text{OPD}_2(p,q)$ is the OPD measured with the surrounding medium that has RI, $n_{m,2}$. Thus, we can isolate the thickness distribution from the first equation, yielding:

$$h(p,q) = \frac{\text{OPD}_1(p,q)}{[n_{cell}(p,q) - n_{m,1}]}, \tag{14}$$

such that by inserting it in the second equation in Eq. (13) and isolating the integral RI, $n_{cell}(p,q)$, we get:

$$n_{cell}(p,q) = \frac{\text{OPD}_2(p,q) \cdot n_{m,1} - \text{OPD}_1(p,q) \cdot n_{m,2}}{\text{OPD}_2(p,q) - \text{OPD}_1(p,q)}, \tag{15}$$

enabling the extraction of the thickness profile as well, by inserting this expression to Eq. (14).

This formulation is valid only if the RI and thickness of the cell both remain unchanged between the two measurements, as would be true for static cells, given that the two mediums have identical osmolarity. Nevertheless, Rappaz et al. [5] have generalized this method to cells with slow dynamics. In the perfusion procedure proposed in this paper, the solution exchange of the perfusion chamber takes 30 [sec], a time during which living cells present micromovements, resulting, for each pixel, in a cell-mediated temporal phase fluctuations, which produce artifacts in the calculations of $n_{cell}(p,q)$ and $h(p,q)$. To solve this, it was proposed to calculate the mean integral refractive index of the cell by taking the average of the integral refractive index over a cellular surface determined by a process involving a gradient-based edge detection algorithm and an erosion procedure, allowing to remove peripheral pixels with low signal to noise ratio, from which a consistent calculation of $n_{cell}(p,q)$ is not possible. Such a mean integral refractive index value presents a higher temporal stability than the full distribution $n_{cell}(p,q)$, and can be used to extract the thickness profile from Eq. (14) to gain more accurate results [5,33].

Jafarfard et al. [35], Xin et al. [36] and Boss et al. [37] have suggested using the same medium for the two measurements, but with a different illumination wavelength at each, such that the RI of the medium will differ between the two measurements. The great advantage here over the former medium-changing approach is that the two measurements can be done simultaneously, allowing decoupling even for dynamic cells. For this to work, however, the medium has to be very dispersive while the cell cannot be dispersive (to prevent $n_{cell}(p,q)$ from changing), a condition that is not always met. For example, Jafarfard et al. [35] suggested using ethylene glycol (C2H6O2) as a medium, for which the RI differs by 0.02 by altering the wavelength from 532 nm to 632 nm. However, this solution is toxic to cells, and thus may prohibit biologically relevant studies and possibly also dynamic observations. Alternatively, the solution suggested by Boss et al. [37] (fast green FCF dye), is both highly dispersive and an FDA approved food dye which presents relative low toxicity for living animals, and thus may be more suitable for dynamic studies.

In all methods presented above, the extracted integral RI profile is a 2-D matrix, representing in each pixel the thickness-average RI of the cell. Of course, most cells are inhomogeneous with RI values that might also change in the axial dimension. In the next section, we present another group of approaches which is based on tomography, requiring the acquisition of the cell digital holograms from multiple points of view, to retrieve the 3-D, rather than integral 2D, RI distribution.

5. **Extracting the 3-D RI profile by tomographic phase microscopy**

In tomographic phase microscopy (TPM), we acquire digital holograms of a sample from multiple viewing angles. Once all holograms are processed, we map their 2-D Fourier Transforms into the 3-D Fourier space of the object, and thus reconstruct the 3-D RI distribution of the object [38-43, 53].

Acquiring holograms from multiple angles can be done by one of two main approaches: either illumination rotation or sample rotation. In illumination rotation [42, 44-46,49,54], the illumination beam rotates, whereas the specimen and the optical setup remain stationary. Although this approach does not require disturbing the sample during data acquisition, the acceptance angle of the illumination is limited, typically up to $\pm 70°$, causing missing points in the angular spectrum, which need to be interpolated digitally [48]. In sample rotation, on the other hand, the sample itself is rotated. This can be done by either a physical perturbation such as a rotating micropipette [40], rotating patch clamping [41], or rotating an optical fiber/microcapillary consisting of the sample [39,47], by exploiting the random rolling of cells while they are flowing along a microfluidic channel [50,51], or by cell-direct micro-manipulation techniques such as holographic optical tweezers [48] or dielectrophoretic cell rotation [52], where the latter allows a full 360° rotation on either axis, enabling filling the entire spectrum, as explained in Section 5.2. Although the angular coverage of the sample rotation methods is better than the illumination rotation methods, sample rotation is prone to more error originated from the rotation stability. In either approach, TPM is limited to samples with dynamics slower than the angular acquisition rate.

As soon as the angular projections are obtained, they can be processed to the 3-D RI distribution of the cell. There are two basic approaches for the tomographic mapping procedure: optical projection tomography (OPT) and optical diffraction tomography (ODT); The OPT approach neglects diffraction and assumes that light propagates along straight lines with unchanged spatial frequency vectors, a good approximation for objects with a small RI span, such as weakly scattering biological cells in vitro [40-43], while ODT takes diffraction of light into account, and more accurately calculates the 3-D RI profile of the imaged cell [44-46, 55].

### 5.1 Optical projection tomography

Optical projection tomography neglects diffraction. Under this assumption, the OPD expression in Eq. (2) (and its digital counterpart in Eq. (3)) can be directly used. The Radon algorithm for tomographic reconstruction as well as the entropy-based algorithms both rely on this simple formulation to retrieve the 3-D RI distribution $n(p,q,l)$.

#### 5.1.1 Radon algorithm

Eq. (2) states that the OPD is the integration of the difference between the RI of the cell and the medium in the direction of light propagation $z$; thus, the OPD is the Radon transform of the function inside the brackets in Eq. (2) [38]. According to this assumption, in order to reconstruct the 3-D object $n(p,q,l) - n_m$, we can employ the Fourier slice theorem, mapping the 2-D Fourier transform of each OPD map taken at angle $\theta$ into a radial plane rotated at an angle $\theta$ around $k_{q_0}$ in the 3-D spatial Fourier space $(k_{p_0}, k_{q_0}, k_{l_0})$ of the object (see Fig. 3). The 3-D RI distribution of the original object can then be obtained by performing an inverse 3-D Fourier Transform, and adding the value of the RI of the medium, $n_m$. According to this formulation, the 2-D projections are mapped to the Fourier space as radial planes; thus 180° rotation on a single axis allows filling the entire Fourier-space.

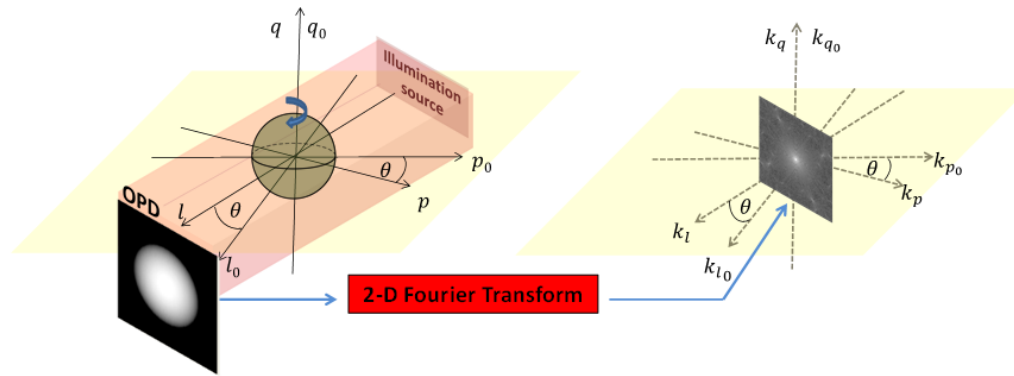

**Fig 3:** The Fourier slice theorem maps the Fourier transform of a 2-D projection into a radial plane in the 3-D Fourier space. $(p_0, q_0, l_0)$ and $(k_{p_0}, k_{q_0}, k_{l_0})$ are stationary coordinate systems in the image and Fourier spaces, respectively. $(p, q, l)$ and $(k_p, k_q, k_l)$ are coordinate systems rotated around the $q_0$ and $k_{q_0}$ axes, respectively, at angle $\theta$. The cell rotation direction is indicated by the arrow around the $q_0$ axis.

#### 5.1.2 Entropy algorithm

In this method, suggested by Xin et al., the sample is imaged only in two orthogonal directions, and two linear equations are produced for each voxel using Eq. (3), such that the equations have infinite solutions theoretically. In order to find the optimal solution, the method of maximum entropy is adopted, turning it into an optimization problem that can be solved iteratively [36, 66].

*5.2 Optical diffraction tomography*

The diffraction tomography algorithm, introduced by Wolf at 1969 [67], takes diffraction of light into account. The diffraction algorithm can be implemented using either the Born or Rytov approximations [38,43,68], both of which take the incident field as the driving field at each point of the scatterer and consider the scattered electrical field due to the presence of a sample [46]. While the Born approximation is valid for a weak scattering field from a relatively small sample, and only works for large samples given a low RI contrast, the Rytov approximation is valid as long as the phase gradient in the sample is small and is independent of the sample size, and is therefore usually preferred over the Born approximation [43,46,68]. When applying the optical diffraction algorithm using the Rytov approximation, the Rytov field for each hologram is calculated as [43]:

$$u_{Rytov}(p,q) = \log\left(\frac{|E_s(p,q)|}{|E_0(p,q)|}\right) + i \cdot \varphi(p,q), \tag{16}$$

where $|E_s(p,q)|$ is the amplitude extracted from the sample hologram, $|E_0(p,q)|$ is the amplitude extracted of a sample-free acquisition taken for reference, $\varphi(p,q)$ is the phase profile extracted from the sample hologram, and $i$ is the imaginary unit. The 2-D Fourier transform of the Rytov field for each hologram taken at angle $\theta$ is then mapped into a 2-D hemispheric surface (Ewald sphere) rotated at an angle $\theta$ around $k_{q_0}$ in the 3-D spatial Fourier space $(k_{p_0}, k_{q_0}, k_{l_0})$ (see Fig. 4). The 3-D RI distribution of the original object can then be obtained by performing an inverse 3-D Fourier Transform, and applying Eq. (17) [43]:

$$n(p,q,l) = \sqrt{n_m^2 - \frac{\lambda}{2\pi} \cdot f(p,q,l)}, \tag{17}$$

where $n_m$ is the RI of the medium, $\lambda$ is the illumination wavelength and $f(p,q,l)$ is the result of the inverse 3-D Fourier transform.

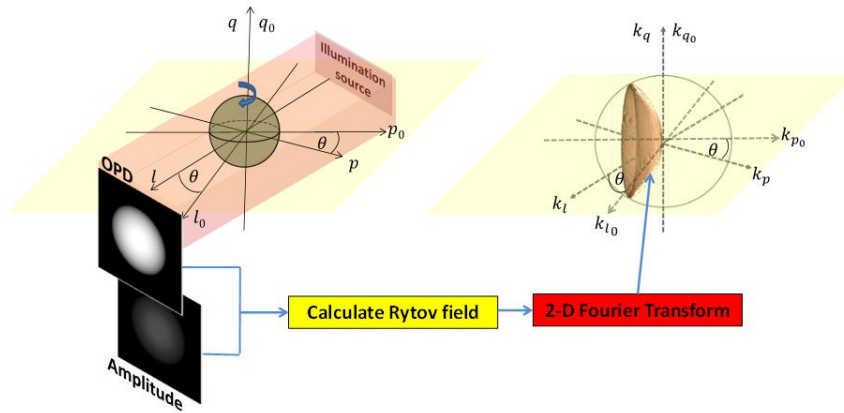

Fig 4: The optical diffraction tomography algorithm maps the Fourier transform of the Rytov field of a 2-D projection into an Ewald sphere in the 3-D Fourier space. $(p_0, q_0, l_0)$ and $(k_{p_0}, k_{q_0}, k_{l_0})$ are stationary coordinate systems in the image and Fourier

spaces, respectively. $(p, q, l)$ and $(k_p, k_q, k_l)$ are coordinate systems rotated around the $q_0$ and $k_{q_0}$ axes, respectively, at angle $\theta$. The cell rotation direction is indicated by the arrow around the $q_0$ axis.

According to this formulation, the 2-D projections are mapped to the Fourier space as hemispheric surfaces; thus 180° rotation on single axis is not sufficient to fill the entire Fourier space. In fact, even a full 360° rotation on a single axis does not allow filling the entire Fourier-space, as this leaves missing data near the axis of rotation (namely the missing cone problem), possibly leading to artifacts and asymmetric resolution [53]. Only a full 360° rotation on one axis combined with an additional rotation around the orthogonal axis allows filling the missing cone, supplying a full coverage of the Fourier space.

An example of applying tomographic phase microscopy to a yeast cell (Saccharomyces Cerevisiae, longitudinal diameter range of 5–10 μm) in suspension during reproduction by budding is given in Fig. 5. In this example, taken from Ref. [48], the cell was rotated in 180° with 5° increment using holographic optical tweezers, and reconstructed by both OPT (Figs. 5(a) and 5(b)) and ODT (Figs. 5(c) and 5(d)).

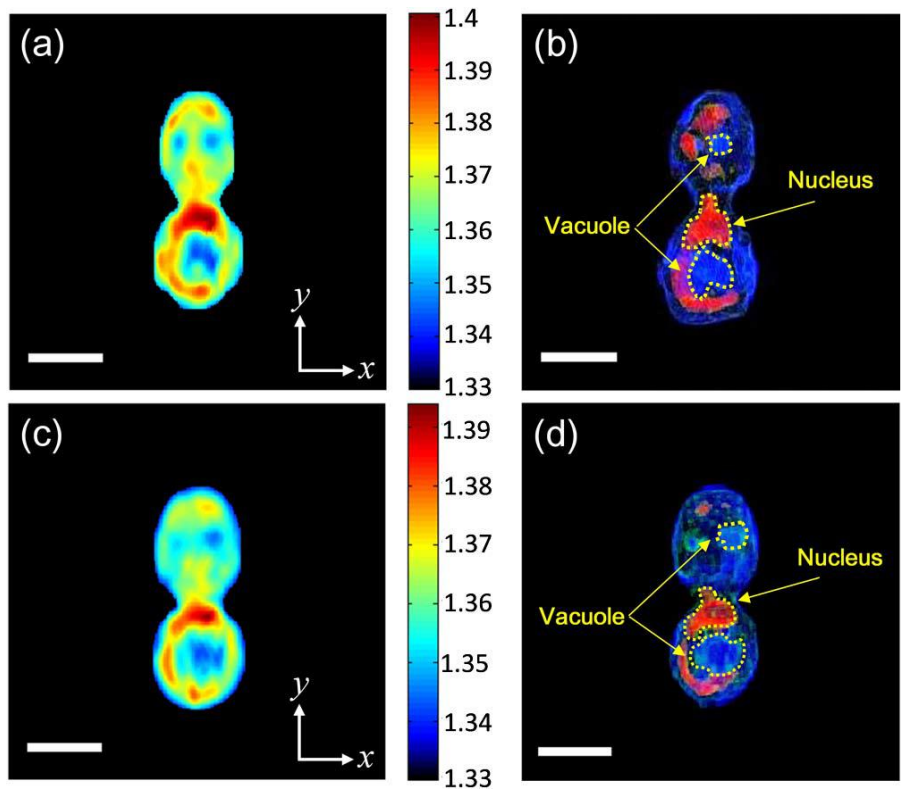

Fig 5: TPM-based three-dimensional refractive index map of yeast cells obtained by (a), (b) the Radon algorithm (OPT), and (c), (d) the optical diffraction method. (a), (c) Central Z slice. (b), (d) Volumetric renderings. The background refractive index is n = 1.33010 ± 0.00047. Note, however, that the background is not a part of the TPM, since it is not rotated with the cells. The white scale bar represents 5 μm upon the sample [48].

More examples of the 3-D RI obtained by using TPM can be seen in Refs. [39-42, 44-52, 54-56].

6. **Medical and biological applications for the RI measurement**

RI measurements give insight to the structure and function of the cell for both medical diagnosis and biological assays.

Numerous recent DHM studies have overturned the former consensus that the nucleus is more dense than the cytoplasm in light of its biological function, by proving that the RI of the cell nucleus is in fact lower than that of cytoplasm [26-28, 42].

RI measurements can also be used as a tool for developing new medications; Ekpenyong et al. [24] have recently shown that bacterial infection of macrophages induces decrease in refractive index, enabling using RI as a parameter for monitoring host-pathogen interactions at a single cell level, providing a fast and reliable assay for evaluating new antibiotics in the context of increasing Salmonella resistance to extant antimicrobial treatments.

RI measurements are also able to differentiate normal cells from pathological cells, as has been demonstrated for numerous diagnostic clinical applications; by obtaining the RI of red blood cells, the concentration of hemoglobin they carry can be determined, enabling diagnosing hypochromic anemia [69]. RI is also expected to be a good criterion when choosing sperm cells for in vitro fertilization [30], as it indicates the density of the genetic material. In addition, Choi et al. [32] have demonstrated RI-based identification of live cancer cells by proving that cancer cells are associated with higher percentage of large RI distribution, stating that the RI assessment of a cell might be a key indicator to efficiently discriminate the cell malignancy. This agrees well with previous results [70], and may be explained by the fact that compared to normal cells, cancer cells have more protein stored in the relatively larger nucleus in order to adapt to the rapid cell division [71]. Other than the specific value of the RI as a probe for pathology, the 3-D RI distribution produced by TPM measurements allows classification of pathologies that alter the morphology of the cell, such as malaria [44, 46] and thalassemia [51].

7. Conclusions

This manuscript reviewed the state-of-the-art approaches for extracting the cellular RI from DHM measurements by solving the RI-thickness coupling problem in label-free DHM. According to this problem, the measurement provided by DHM yields the product of the thickness-average RI and physical thickness of the cells (Section 2). Each of the RI extraction methods reviewed in this paper has its weaknesses and strengths, which should be considered when choosing a decoupling approach. We have generally reviewed two approaches for extracting the RI: obtaining the thickness of the cell and using it to extract the RI, or decouple thickness from RI by multiple DHM measurements.

Regarding the first approach of obtaining the thickness, several methods have been presented. While evaluating the thickness profile by assuming a spherical model (Section 3.1) is undoubtedly the simplest and quickest of all methods, and can easily be applied to dynamic samples, such as cells during fast flow, it may not be accurate enough for many applications, and is not relevant for attached cells, as well as for all inherently non-spherical cells (such as neurons or sperm cells). Even though an ellipsoid model makes the approach slightly more general, it is still most accurate when applied to perfect spheres, as the evaluation of the third radius of the ellipsoid as the average may not be correct for certain applications.

Measuring the physical thickness distribution by another measurement method (Section 3.2) is not always practical as it requires imaging each cell by two modalities, sometimes on two different systems, and AFM, confocal microscopy, and optical coherence microscopy are all time consuming even when performed simultaneously with DHM, making it unsuitable for highly dynamic samples. Nevertheless, by performing this approach on a large population of cells, RI statistics can be collected and used for future extraction of geometrical properties. Confining the cell to a known thickness (Section 3.3), though yielding efficient RI measurements that can be done automatically and in high throughput during flow in a micro-channel, disturbs the cell and might result in losing morphological information.

Regarding the second approach of multiple measurements, we have presented several methods as well. First, as presented in Section 4, dual DHM measurements allow evaluating both the local thickness and integral RI simultaneously without the need of an additional modality, yet it too is limited to samples that do not change between the measurements, and requires a specialized setup allowing for medium replacement. If only the illumination wavelength is replaced between measurements, it must be verified that the medium is dispersive and the cell is not; since most cells in vitro have negligible RI dependency in the visible wavelength range, this is a reasonable assumption that can be met by choosing an appropriate suspension medium. The great advantage of this method over the former is that the two measurements can be taken simultaneously, allowing decoupling even for highly dynamic cells.

All methods mentioned above in this section yield an integral 2-D RI distribution rather than a 3-D RI distribution. The fact that the RI extracted is an average over the axial axis means that it may blur out important details, including the cellular organelles. TPM, presented in Section 5, solves this problem by acquiring many DHM measurements from different viewing angles on the cell. While TPM clearly gives the most elaborate information, yielding a 3-D RI distribution and morphology, it is more difficult to implement due to the complexity of the angular scanning optical setup and is currently not suitable for imaging samples with fast dynamics.

Note that DHM is also called quantitative phase microscopy (QPM) [72, 73] and interferometric phase microscopy (IPM) [7, 74, 75]. Unlike DHM, the terms IPM and QPM specifically refer to cases when only the phase (rather than phase and amplitude) is of interest. In addition, the term QPM may refer to cases where the phase was acquired by non-interferometric methods, such as transport of intensity [76]. While this review addresses the mathematical formulations regarding the phase difference acquired by interferometric methods, all decoupling methods presented can be generalized to all quantitative phase acquisition methods, even the non-interferometric ones.

**Funding**

This work was supported by the Horizon 2020 European Research Council (ERC) [grant number 678316]; Tel Aviv University Center for Light-Matter Interaction.